\documentclass[twocolumn,showkeys,aps,prb,showpacs]{revtex4-1}
\UseRawInputEncoding
\usepackage{graphicx}
\usepackage[CJKbookmarks,dvipdfm,colorlinks,linkcolor=red,citecolor=blue]{hyperref}

\begin{document}

\title{Strain effects on topological and valley properties  of Janus monolayer $\mathrm{VSiGeN_4}$}

\author{San-Dong Guo$^{1}$, Wen-Qi Mu$^{1}$,  Jia-Hao Wang$^{1}$, Yu-Xuan Yang$^{1}$, Bing Wang$^{2}$ and Yee-Sin Ang$^{3}$}
\affiliation{$^1$School of Electronic Engineering, Xi'an University of Posts and Telecommunications, Xi'an 710121, China}
\affiliation{$^2$Institute for Computational Materials Science, School of Physics and Electronics,Henan University, 475004, Kaifeng, China}
\affiliation{$^3$Science, Mathematics and Technology (SMT), Singapore University of Technology and Design (SUTD), 8 Somapah Road, Singapore 487372, Singapore}
\begin{abstract}
Strain is an effective method to tune the electronic properties of two-dimension (2D) materials, and can induce  novel phase transition.  Recently, 2D $\mathrm{MA_2Z_4}$ family materials are of interest because of their  emerging topological,
magnetic and superconducting properties. Here, we investigate the impact of strain effects ($a/a_0$:0.96$\sim$1.04)
on the physical properties of Janus monolayer $\mathrm{VSiGeN_4}$ as a  derivative  of $\mathrm{VSi_2N_4}$ or $\mathrm{VGe_2N_4}$,  which possesses dynamical, mechanical and thermal  stabilities.
For out-of-plane magnetic anisotropy, with increasing strain, $\mathrm{VSiGeN_4}$ undergoes transition between  ferrovalley semiconductor (FVS), half-valley-metal (HVM), valley-polarized quantum anomalous Hall insulator (VQAHI), HVM and FVS. These imply twice topological phase transitions,  which are related  with
sign-reversible  Berry curvature  and  band inversion  between $d_{xy}$+$d_{x^2-y^2}$ and $d_{z^2}$ orbitals for K or -K valley.
The  band inversion also leads to transformation of valley splitting strength between valence and conduction bands.
However, for in-plane magnetic anisotropy, no special quantum anomalous Hall (QAH) states and  valley polarization exist within the considered strain range.
  The  actual magnetic anisotropy energy (MAE) shows no special QAH and HVM states in monolayer $\mathrm{VSiGeN_4}$. Fortunately, these can be easily  achieved by external magnetic field, which  adjusts the easy magnetization axis of  $\mathrm{VSiGeN_4}$ from in-plane one to out-of-plane one.  Our findings shed light on how strain can be employed to engineer the electronic states of  $\mathrm{VSiGeN_4}$, which may open new perspectives for
 multifunctional quantum devices in valleytronics and spintronics.

\end{abstract}
\keywords{Strain, Magnetic anisotropy energy,  Phase transition ~~~~~~~~~~~~~Email:sandongyuwang@163.com}

\maketitle

\section{Introduction}
Magnetism of 2D systems  is one of the most fascinating
properties of material due to its interplay with the
other important properties of materials such as superconductivity, ferrovalley (FV), ferroelectricity, piezoelectricity and QAH effects.
However, based on Mermin-Wagner theorem, long-range magnetic order is prohibited in a 2D system\cite{a1}.
Fortunately,   2D intrinsic long-range ferromagnetic
(FM) order  semiconductors, $\mathrm{Cr_2Ge_2Te_6}$ and
$\mathrm{CrI_3}$,  have been achieved experimentally, obtained from
their van der Waals (vdW) layered bulk materials\cite{a2,a3}, due to the stabilization of  FM order  by magnetic anisotropy.
In addition to this,  the direction of  magnetic anisotropy has  important influence on the topological and valley properties of some 2D materials, because it can affect the symmetry of such 2D systems\cite{a4,a5,a6,a7}. For example in monolayer $\mathrm{RuBr_2}$, FV to HVM  to QAH to HVM to FV transitions can be induced by increasing  the electron correlation $U$  with a fixed out-of-plane magnetic anisotropy, but  no special QAH states and  valley polarization can be observed for the in-plane case\cite{a7}.
Thus, it may be a very interesting  to tune the magnetic anisotropy of 2D systems by external field, such as biaxial strain, electric field, and correlation effects.

Strain engineering is an important strategy for tuning the electronic,  topological, thermoelectric, piezoelectric and magnetic properties of 2D materials, which has been widely used in the modulation of physical and chemical properties\cite{a8}.  The QAH state in the $\mathrm{VN_2X_2Y_2}$ nanosheets (X=B-Ga, Y=O-Te) can be induced by
strain, and the valley polarization can also be switched from the bottom conduction band to the top valence band\cite{a9}. For monolayer $\mathrm{MBr_2}$ (M=Ru and Os),  compressive strain can induce phase transitions in the materials from FVS
to HVM to VQAHI to HVM to FVS\cite{a10}. However, in these works, the intrinsic MAE as a function of strain has not been considered, and out-of-plane magnetic anisotropy is assumed to be fixed within the considered strain range. Our recent works show that an increasing strain can induce  switching of the magnetic anisotropy from out-of-plane one to in-plane one\cite{a7}, thus producing  manifold electronic states. Thus, strain engineering may produce  complex phase transition of electronic states by tuning the magnetic anisotropy.
\begin{figure*}
  \includegraphics[width=13cm]{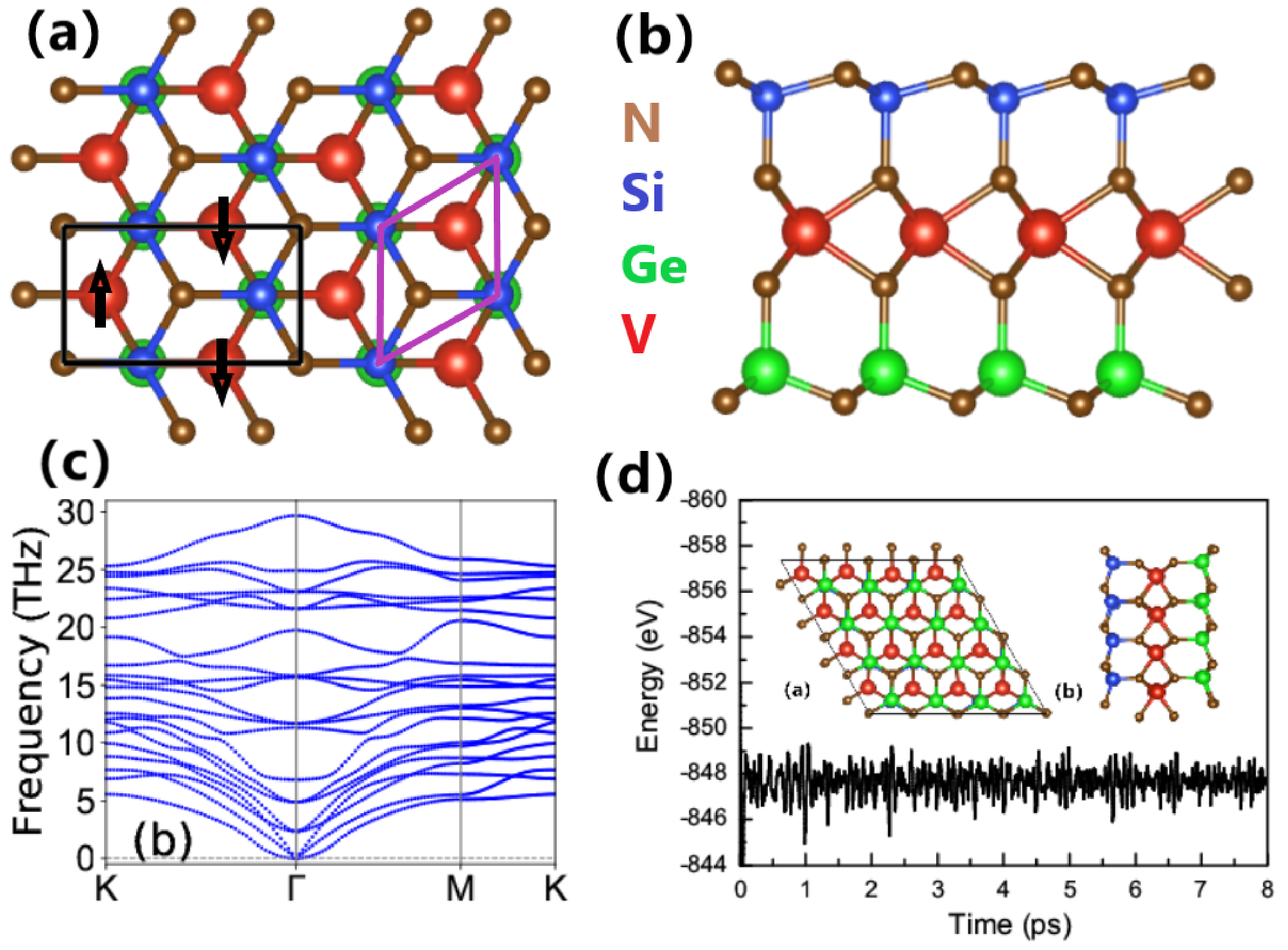}
  \caption{(Color online)For $\mathrm{VSiGeN_4}$ monolayer,  (a): top view and (b): side view of  crystal structure. The primitive (rectangle supercell) cell is
   shown by purple (black) lines, and the AFM configuration is marked with  black arrows in (a). (c):the phonon dispersion curves. (d):the  total energy fluctuations as a function of simulation time at 300 K, and insets show the
 final structures (top view (a) and side view (b)) of $\mathrm{VSiGeN_4}$ after 8 ps at 300 K.  }\label{st}
\end{figure*}

\begin{figure}
  \includegraphics[width=8cm]{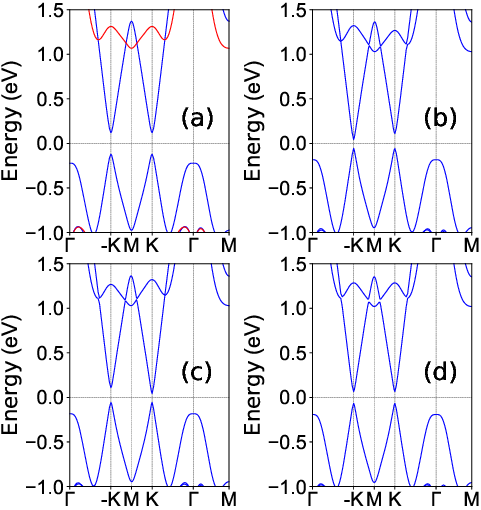}
  \caption{(Color online) Energy  band structures of  $\mathrm{VSiGeN_4}$ (a) without SOC; (b), (c) and (d) with SOC for magnetic moment of V along the positive $z$, negative $z$, and positive $x$ direction, respectively.  In (a), the blue (red) lines represent the band structure in the spin-up (spin-down) direction. }\label{band-z}
\end{figure}

\begin{figure*}
  \includegraphics[width=16cm]{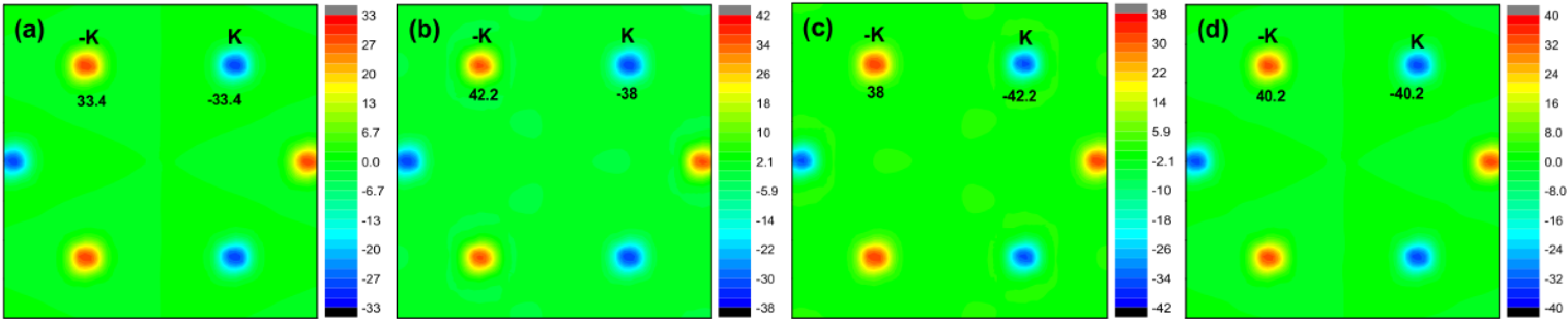}
  \caption{(Color online)For  $\mathrm{VSiGeN_4}$ monolayer, the corresponding Berry curvature distribution in 2D BZ (a) without SOC; (b), (c) and (d) with SOC for magnetic moment of V along the positive $z$, negative $z$, and positive $x$ direction, respectively.}\label{s-b0}
\end{figure*}

\begin{figure}
  \includegraphics[width=8cm]{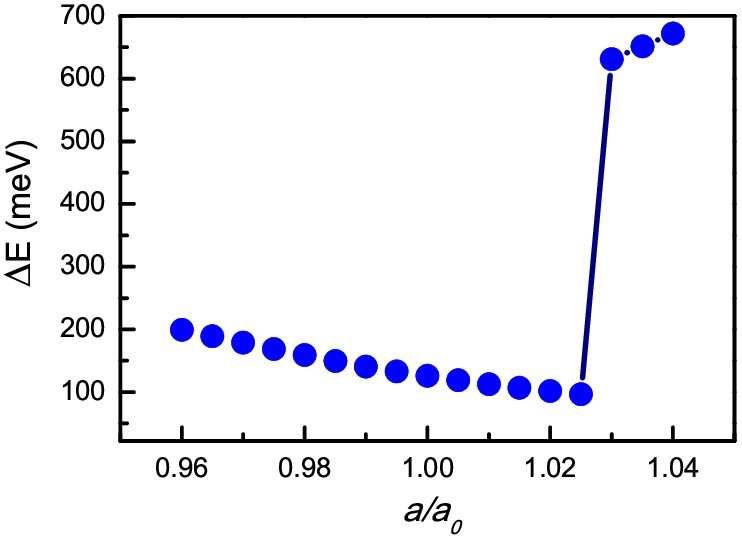}
  \caption{(Color online)For  $\mathrm{VSiGeN_4}$ monolayer, the  energy differences $\Delta E$ between  AFM and FM ordering  as a function of $a/a_0$.}\label{s-e}
\end{figure}

In 2020, the septuple-atomic-layer
2D $\mathrm{MoSi_2N_4}$ and $\mathrm{WSi_2N_4}$ have been  successfully
synthesized  by the chemical vapor deposition
method\cite{a11}. Subsequently,  2D $\mathrm{MA_2Z_4}$
family  with a septuple-atomic-layer structure  has been constructed by intercalating a $\mathrm{MoS_2}$-type
monolayer $\mathrm{MZ_2}$ into an InSe-type monolayer $\mathrm{A_2Z_2}$, and the family possesses  emerging topological,
magnetic, valley, superconducting and electrical contact properties\cite{a12,a12-1,a12-2}. Then,  Janus 2D materials in the new  2D $\mathrm{MA_2Z_4}$ family are  proposed, such as $\mathrm{MSiGeN_4}$ (M=Mo and W) and $\mathrm{SrAlGaSe_4}$, and some novel properties can be achieved in these Janus materials, such as Rashba spin splitting and out-of-plane piezoelectric polarizations\cite{a13,a14}. Recently,  Janus $\mathrm{VSiGeN_4}$ monolayer  is predicted to be  a thermodynamically stable
intrinsic  2D ferromagnet\cite{a15}.

In this work, we investigate strain effects on topological and valley properties of  Janus $\mathrm{VSiGeN_4}$ monolayer,  and reveal the
importance of magnetic anisotropy in determining its magnetic, topological and valley properties.
It is found that different strain strengths can drive the system into
different novel electronic states (FVS, HVM and VQAHI) with fixed out-of-plane case, enabling a rich phase diagram.
However, for in-plane case, only common  magnetic states appear.  Due to  weak spin-orbit
coupling (SOC) in $\mathrm{VSiGeN_4}$, the magnetic shape anisotropy (MSA) induced by the magnetic dipolar interaction can overcome the magnetocrystalline anisotropy (MCA) to evince an easy-plane in considered strain range. So, strained $\mathrm{VSiGeN_4}$ is intrinsically a common magnetic semiconductor. However, these topological and valley states can be achieved by small external magnetic field.
With increasing $a/a_0$,   the MCA energy  firstly switches from in-plane to out-of-plane.   Further increasing  $a/a_0$ will drive two additional transitions in the
 MCA from  out-of-plane to in-plane to out-of-plane. Several transitions in the MCA  are further identified by calculating MCA versus $U$.
Our works  highlight the role of
magnetic anisotropy for $\mathrm{VSiGeN_4}$,
and deepen our understanding of strain along with magnetic anisotropy induced
topological and valley states.

The rest of the paper is organized as follows. In the next
section, we shall give our computational details and methods.
 In  the next few sections,  we shall present structure and stabilities, electronic states  and strain effects on  physical properties  of  $\mathrm{VSiGeN_4}$ monolayer. Finally, we shall give our discussion and conclusion.

\section{Computational detail}
Within density-functional
theory  (DFT)\cite{1},  we perform spin-polarized  first-principles calculations  by employing the projected
augmented wave method,  as implemented in VASP code\cite{pv1,pv2,pv3}.
The generalized gradient approximation of Perdew-Burke-Ernzerhof (PBE-GGA)\cite{pbe} is adopted as exchange-correlation functional.
The energy cut-off of 500 eV,  total energy  convergence criterion of  $10^{-8}$ eV and  force
convergence criteria of less than 0.0001 $\mathrm{eV.{\AA}^{-1}}$ on each atom   are used to attain accurate results.
 A vacuum space of more than 30 $\mathrm{{\AA}}$ is used to avoid the interactions
between the neighboring slabs.
The $\Gamma$-centered 16$\times$16$\times$1 k-point meshs are sampled in  the Brillouin zone (BZ) for structure optimization, electronic structures and elastic stiffness tensor, and 9$\times$16$\times$1 Monkhorst-Pack k-point meshs for FM/antiferromagnetic (AFM)  energy  with rectangle supercell.
The on-site Coulomb correlation of V  atoms is considered by using  GGA+$U$  method within  the rotationally invariant approach proposed by Dudarev et al\cite{u}, and the $U$$=$3.2 eV is used, which
has been also used in ref.\cite{a15}. The SOC effect is explicitly included to investigate MCA, electronic and topological properties of  $\mathrm{VSiGeN_4}$ monolayer.

 The vibrational properties are investigated  by
the finite-displacement method  with a 5$\times$5$\times$1 supercell, as  implemented in the
Phonopy code\cite{pv5}. We use strain-stress relationship (SSR)  to attain  elastic stiffness tensor  $C_{ij}$, and
the   2D elastic  coefficients $C^{2D}_{ij}$ have been renormalized by   $C^{2D}_{ij}$=$L_z$$C^{3D}_{ij}$, where  the $L_z$  is  the cell height along $z$ direction.
The Berry curvatures  are calculated  directly from  wave functions  based on Fukui's
method\cite{bm}, as implemented in VASPBERRY code\cite{bm1,bm2}.
The mostly localized Wannier functions including the $d$-orbitals of V atom and the $p$-orbitals of Si, Ge and N atoms are constructed on a k-mesh of 16$\times$16$\times$1, and then are used to calculate edge states  using  Wannier90 and  WannierTools packages\cite{w1,w2}.
The  energy band structures of $\mathrm{VSiGeN_4}$  calculated by DFT and fitted  by
Wannier90 at $a/a_0$=0.993 (in topological state) are plotted in FIG.1 of electronic supplementary information (ESI), which  confirms the fitting accuracy.

\begin{figure}
   \includegraphics[width=7cm]{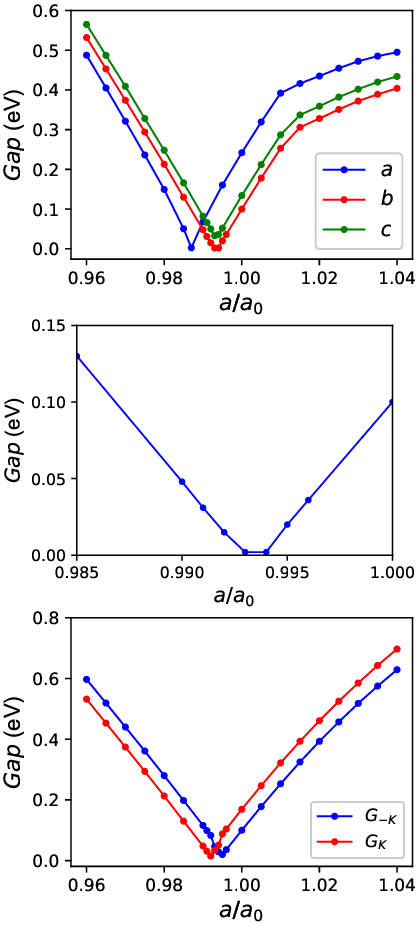}
  \caption{(Color online)Top panel:  the  global energy band gap without SOC (a) and with SOC [out-of-plane  magnetic anisotropy (b) and  in-plane  magnetic anisotropy (c)]; Middle panel: the enlarged view of  global energy band gap  with SOC [out-of-plane  magnetic anisotropy] near $a/a_0$=0.993; Bottom panel: the energy  band gaps for -K and K valleys as a function of   $a/a_0$ with SOC [out-of-plane  magnetic anisotropy]. }\label{s-gap}
\end{figure}
\begin{figure*}
  \includegraphics[width=16cm]{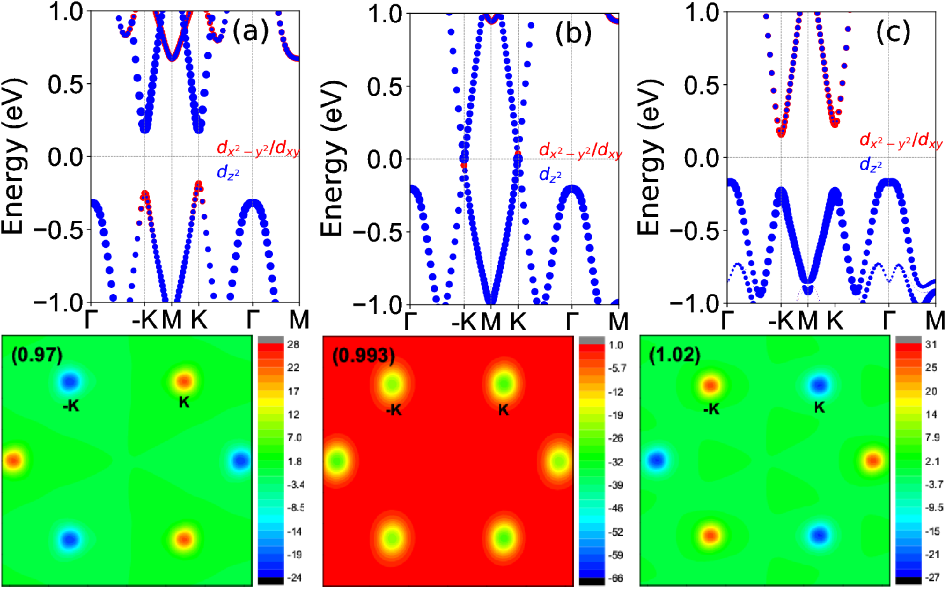}\\
\caption{(Color online)For $\mathrm{VSiGeN_4}$ monolayer with out-of-plane  magnetic anisotropy, the V-$d_{x^2-y^2}$+$d_{xy}$ and $d_{z^2}$-orbital characters energy band structures at representative  $a/a_0$$=$0.97 (a), 0.993 (b) and 1.02 (c), along with the corresponding Berry curvature distribution in 2D BZ.}\label{s-sb}
\end{figure*}

\section{Structure and stabilities}
As shown in \autoref{st}, this structure of $\mathrm{VSiGeN_4}$ monolayer is stacked
by seven atomic layers of N-Si-N-V-N-Ge-N.  This can be regarded
as a sandwich structure, and the middle $\mathrm{VN_2}$ layer is sandwiched by
SiN and GeN bilayers, which can be considered as a Janus structure.
The $\mathrm{VSiGeN_4}$ monolayer can be built  by replacing the Si/Ge atoms of one of two SiN/GeN bilayers in   $\mathrm{VSi_2N_4}$/ $\mathrm{VGe_2N_4}$ monolayer with Ge/N atoms. The symmetry of $\mathrm{VSiGeN_4}$ (No.156) is lower than that of  $\mathrm{VSi_2N_4}$/ $\mathrm{VGe_2N_4}$ (No.187) due to
the lack of  the reflection symmetry with respect to the middle $\mathrm{VN_2}$ layer.  The rhombus primitive cell  and the rectangle supercell are plotted in \autoref{st} (a) along with AFM configuration, and  the first BZ with high-symmetry points is shown in FIG.2 of ESI.
The optimized  lattice constants $a$ of $\mathrm{VSiGeN_4}$ monolayer is 2.959 $\mathrm{{\AA}}$ with FM ordering, which agrees well with previous theoretical value (2.97$\mathrm{{\AA}}$)\cite{a15}.

Our calculations show that $\mathrm{VSiGeN_4}$ stabilizes  into a FM ground state, and the FM state
is  125.7 meV  lower in energy than its AFM state with rectangle supercell. The MAE includes two main terms\cite{re1,re2,re3,re4}: (1) MCA energy ($E_{MCA}$), which is  induced by the SOC, and (2) MSA energy ($E_{MSA}$), which is due to the dipole-dipole (D-D) interaction:
  \begin{equation}\label{d-d}
E_{D-D}=-\frac{1}{2}\frac{\mu_0}{4\pi}\sum_{i\neq j}\frac{1}{r_{ij}^3}[\vec{M_i}\cdot\vec{M_j}-\frac{3}{r_{ij}^2}(\vec{M_i}\cdot\vec{r_{ij}})(\vec{M_j}\cdot\vec{r_{ij}})]
 \end{equation}
where the $\vec{M_i}$ represents the local magnetic moments and $\vec{r_{ij}}$
are vectors that connect the sites $i$ and $j$.
  The $E_{MCA}$  is calculated from a  energy difference between in-plane magnetization
and out-of-plane magnetic anisotropy within SOC. The calculated $E_{MCA}$ of $\mathrm{VSiGeN_4}$ is only -3 $\mathrm{\mu eV}$. For most materials, the magnetic D-D interaction is small compared with the MCA interaction. However, for $\mathrm{VSiGeN_4}$, it may play an
important role due to very  small  $E_{MCA}$. According to \autoref{d-d},  the  $E_{MSA}$ is calculated from a  energy difference with the magnetization rotating from the in-plane
direction to the out-of-plane direction. The calculated $E_{MSA}$ is -17 $\mathrm{\mu eV}$, which dominates the MAE (-20 $\mathrm{\mu eV}$).
The  positive/negative  MAE means
that the easy magnetization axis is perpendicular/parallel to the plane of monolayer. The calculated MAE indicates  in-plane easy magnetization, which means that
there is no energetic barrier to the rotation of magnetization in the $xy$ plane\cite{re5}. So, the $\mathrm{VSiGeN_4}$ can be
considered as a 2D $XY$ magnet\cite{re5,re5-1}.   For a 2D XY magnet with a typical triangle lattice structure,
 a Berezinskii-Kosterlitz-Thouless magnetic transition to a quasi-long-range
phase will produce at a critical temperature. The Monte Carlo simulations have predicted  the critical temperature $T_C=1.335\frac{J}{K_B}$\cite{re6,re7}, where $J$ is the nearest-neighboring exchange parameter and $K_B$ is the Boltzmann
constant.  The $J$  is determined from the energy
difference between  AFM  ($E_{AFM}$) and FM ($E_{FM}$).
 Based on the FM
and AFM configurations, the  AFM and FM energies   can be
obtained by equations:
\begin{equation}\label{pe0-1-2}
E_{FM}=E_0-(6J+2A)S^2
 \end{equation}
  \begin{equation}\label{pe0-1-3}
E_{AFM}=E_0+(2J-2A)S^2
 \end{equation}
where $E_0$ is the total energy of systems without magnetic coupling, and $A$ describes the easy-axis single-ion anisotropy.
The  corresponding $J$ can be attained:
  \begin{equation}\label{pe0-1-3}
J=\frac{E_{AFM}-E_{FM}}{8S^2}
 \end{equation}
The calculated $J$ is 31.43 meV ($S=\frac{1}{2}$), and the $T_C$  is estimated to be  487 K.

The dynamical stability of $\mathrm{VSiGeN_4}$ is verified by its phonon band dispersion,  which is presented in \autoref{st} (c).
Phonon branches show no  imaginary frequencies,  indicating the dynamical stability of $\mathrm{VSiGeN_4}$.
Ab initio molecular
dynamics (AIMD) simulations are further performed to
examine the thermal stability of $\mathrm{VSiGeN_4}$ on a 4$\times$4$\times$1 supercell with a
Nose thermostat of 300 K and a step time of 1 fs. As shown in \autoref{st} (d), during the 8 ps simulation time, the
energy are fluctuated around the equilibrium values without any
sudden changes with  small distortions  in the final
configurations,  indicating its good thermal
stability. The  $\mathrm{VSiGeN_4}$ has two independent elastic
constants of $C_{11}$ and $C_{12}$.   If they  satisfy  Born criteria of $C_{11}>0$ and  $C_{11}-C_{12}>0$\cite{ela,ela1}, the $\mathrm{VSiGeN_4}$ will be mechanically stable. The calculated two  independent elastic
constants of  $\mathrm{VSiGeN_4}$ are $C_{11}$=434.15 $\mathrm{Nm^{-1}}$ and $C_{12}$=125.39 $\mathrm{Nm^{-1}}$, which  satisfy the  Born  criteria of  mechanical stability, confirming  its mechanical stability.

\section{electronic structures}
The magnetic anisotropy has crucial effects on the electronic states of  2D materials\cite{a4,a5,a6,a7}.  It is well known  that the magnetization is a pseudovector.
And then,  the out-of-plane FM  breaks all
possible vertical mirrors of the system, but  preserves the horizontal mirror symmetry.  The preserved horizontal mirror symmetry allows the spontaneous valley polarization and a nonvanishing Chern
number of a 2D system\cite{a4}. Although the magnetocrystalline direction  of
$\mathrm{VSiGeN_4}$ monolayer is in-plane, this  can be easily regulated into out-of-plane  by external magnetic field due to the very small MAE.

For $\mathrm{VSiGeN_4}$ monolayer, the spin-polarized band structures  by using both GGA+$U$ and GGA+$U$+SOC are shown in \autoref{band-z}.
\autoref{band-z} (a) shows  a distinct
spin splitting  due to the exchange
interaction, and  $\mathrm{VSiGeN_4}$ is  a direct narrow band
gap semiconductor  with gap value  of 0.242 eV.  The  valence band maximum (VBM) and conduction band bottom (CBM) are at K/-K point, which are  provided by the spin-up.
The energies of  -K and K valleys  are degenerate for both conduction and valence bands.
The V-$d$ orbitals  lie  in a trigonal prismatic crystal field environment, and the  $d$ orbitals split
into low-lying $d_z^2$ orbital, $d_{xy}$+$d_{x^2-y^2}$  and
$d_{xz}$+$d_{yz}$ orbitals.
According to  projected band structure in FIG.3 of ESI,
 only  top $d_{z^2}$-dominated valence band in spin-up direction is occupied by one electron.
This is expected to lead to a
 magnetic moment of the 1 $\mu_B$  for each V atom, which conforms to the calculated value of 1.1 $\mu_B$.

When including SOC, the  valley polarization can be induced with  out-of-plane magnetic anisotropy, as shown in  \autoref{band-z} (b).
The valley splitting of bottom  conduction  band   is 67  meV,  while the valley splitting of top valence  band is only 2 meV.
For bottom  conduction  band, the energy of K valley
is higher than that of -K valley.  As plotted in  \autoref{band-z} (c), the valley polarization can  be
switched by reversing the magnetization direction (The energy of -K valley
is higher than one of K valley.).  \autoref{band-z} (b) and (c) show that  the gap value of $\mathrm{VSiGeN_4}$ is about 0.10 eV.
Based on \autoref{band-z} (d), no valley polarization can be observed  with in-plane magnetic anisotropy,  and it is still  a direct band
gap semiconductor (0.134 eV).

FIG.3 of ESI show that the $d_{x^2-y^2}$+$d_{xy}$/$d_z^2$ orbitals  dominate -K and K valleys of bottom  conduction  band/top valence  band, which determines the strength of valley splitting. The intra-atomic interaction  $\hat{H}^0_{SOC}$ from SOC  mainly gives rise to valley polarization, which with out-of-plane magnetization can be expressed as\cite{f6,v2,v3} :
\begin{equation}\label{m1}
\hat{H}^0_{SOC}=\alpha \hat{L}_z
\end{equation}
where   $\hat{L}_z$/$\alpha$ is the orbital angular moment along $z$ direction/coupling strength. The resulting energy of K
or -K valley can be written as:
\begin{equation}\label{m3}
E^\tau=<\phi^\tau|\hat{H}^0_{SOC}|\phi^\tau>
\end{equation}
where $|\phi^\tau>$ (subscript  $\tau=\pm1$ as valley index) means  the orbital
basis for -K or K valley.
If $d_{x^2-y^2}$+$d_{xy}$ orbitals  dominate  -K and K valleys, the valley splitting $|\Delta E|$  can be written as:
\begin{equation}\label{m4}
|\Delta E|=E^{K}-E^{-K}=4\alpha
\end{equation}
If the -K and K valleys are mainly from $d_{z^2}$ orbitals, the valley splitting $|\Delta E|$
 is written as:
\begin{equation}\label{m4}
|\Delta E|=E^{K}-E^{-K}=0
\end{equation}
According to FIG.3 of ESI, the valley splitting of bottom  conduction  band  will be very large, and the valley splitting of top valence  band will be very small, which agree well with our calculated results.
With general magnetization orientation, $\Delta E=4\alpha cos\theta$\cite{v3} ($\theta$=0/90$^{\circ}$ denotes out-of-plane/in-plane direction.) for $d_{x^2-y^2}$+$d_{xy}$-dominated -K/K valley. For in-plane one, the valley splitting of $\mathrm{VSiGeN_4}$ will be zero.

When an in-plane longitudinal electric field $E$ is applied,   Bloch electrons can attain anomalous velocity $\upsilon$, which is associated with Berry curvature $\Omega(k)$:$\upsilon\sim E\times\Omega(k)$\cite{q9}. The calculated Berry curvature of
$\mathrm{VSiGeN_4}$ as a contour map in 2D BZ with and without SOC are plotted in \autoref{s-b0}, and their hot spots  are around -K and K valleys.
The four situations all show that Berry curvatures have opposite
signs around -K and K valleys with equal/unequal magnitudes for  valley-nonpolarized/valley-polarized situation.
 When reversing
the magnetization from  $z$ to $-z$ direction, the signs of Berry curvature at -K and K valleys remain unchanged, but
their magnitudes
 exchange to each
other. When the Fermi level falls between the -K and K
valleys with appropriate electron doping, the  Berry curvature forces
the spin-up carriers of K valley to accumulate on one side of the sample by  an applied in-plane electric field,  giving rise to an anomalous valley Hall effect
(AVHE).   When the magnetization
is reversed,  the spin-down carriers  of  -K valley   move
to another side of the sample due to opposite
Berry curvature compared with  one of K valley.

\begin{figure}
  \includegraphics[width=7cm]{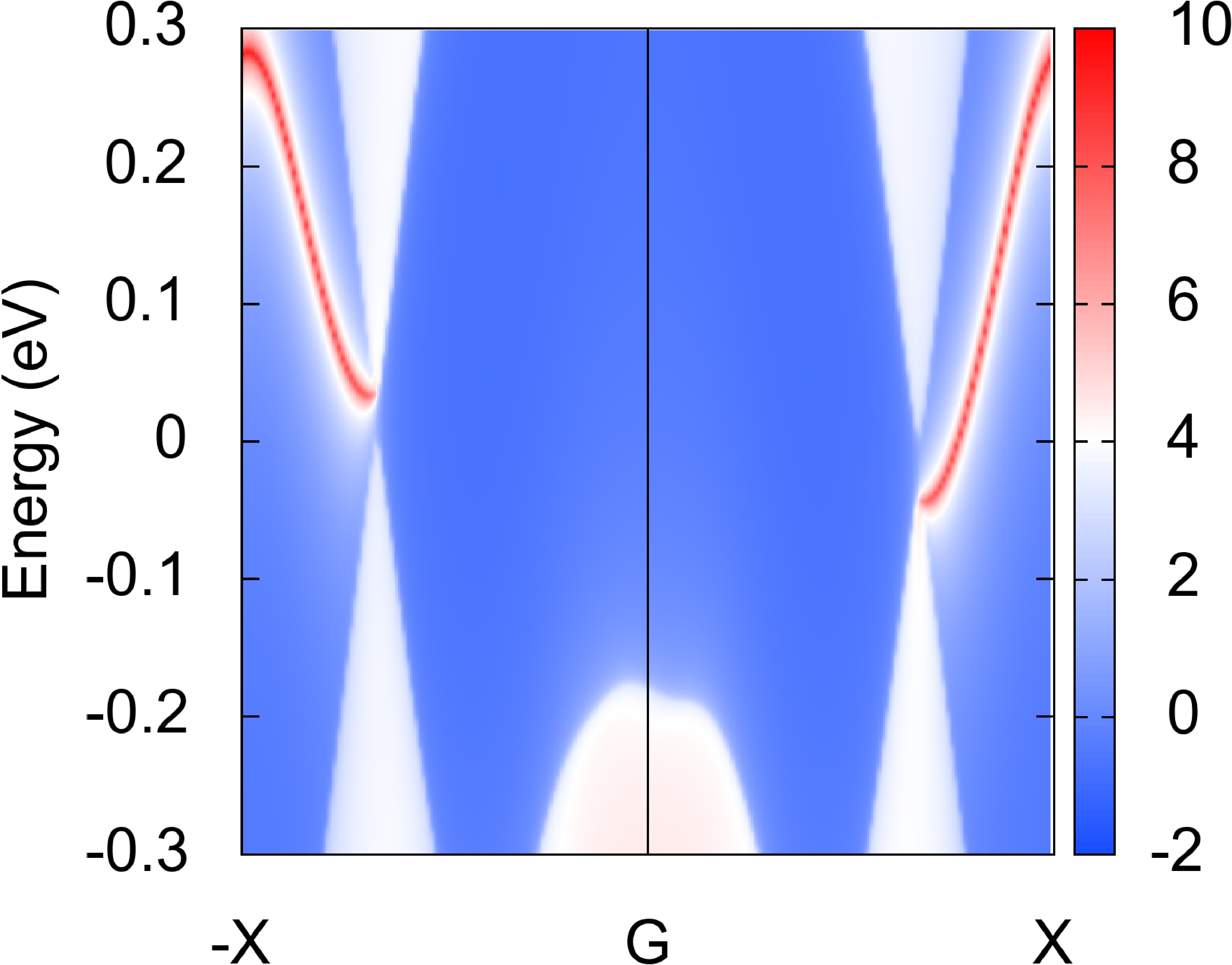}\\
\caption{(Color online)For $\mathrm{VSiGeN_4}$ monolayer with out-of-plane  magnetic anisotropy,  the topological
edge states  at representative  $a/a_0$$=$0.993.}\label{s-s}
\end{figure}

\section{strain effects}
Strain is an effective method to tune the  electronic state of some 2D materials, and can produce novel electronic states, such as FV, QAH and HVM states\cite{a7,a9,a10}.
We use  $a/a_0$  to simulate the biaxial strain, where  $a$/$a_0$ are the strained/unstrained lattice constants.
Here, both  compressive  ($a/a_0$$<$1) and tensile ($a/a_0$$>$1) strains are applied to achieve electronic states tuning($a/a_0$:0.96$\sim$1.04).
As shown in \autoref{s-e}, the
total energy differences between  AFM and FM ordering by using rectangle supercell indicate that the FM state is always the magnetic ground state of $\mathrm{VSiGeN_4}$ in considered strain range. It is found that total energy differences between  AFM and FM ordering at $a/a_0$=1.03 has a sudden jump.
To explain this, the energies of AFM and FM ordering as a function of $a/a_0$ are plotted in FIG.4 of ESI.  Calculated results show that the energy of AFM ordering suddenly increases at $a/a_0$=1.03.  To further reveal the underlying causes, the  magnetic moments of  V atom for both AFM and FM ordering as a function of $a/a_0$ are plotted in FIG.5 of ESI. The energy difference jump is due to an abrupt change of the magnetic moment of  V atom for  AFM ordering, which reduces the magnetic interaction energy.

Next, the strain effects on electronic structures of $\mathrm{VSiGeN_4}$ are investigated. Firstly, the total energy band gaps as a function of $a/a_0$ without SOC are
plotted in \autoref{s-gap}, and energy band structures at some representative $a/a_0$ values are shown in FIG.6 of ESI. When the $a/a_0$ changes from 0.96 to 1.04,  the gap firstly closes at about $a/a_0$=0.987, and then continues  to increase.  Before the energy gap closes, $\mathrm{VSiGeN_4}$ is a direct
gap semiconductor with VBM (CBM) at K/-K point. When $a/a_0$$>$0.987, $\mathrm{VSiGeN_4}$ is still a direct gap semiconductor at small $a/a_0$.
When $a/a_0$$>$1.01, $\mathrm{VSiGeN_4}$ is  an indirect
gap semiconductor. The CBM is at
the K/-K point, whereas the VBM
 deviates  slightly from $\Gamma$ point. In considered strain range, the K and -K valleys   are always provided by the spin-up.

When including SOC, the magnetic anisotropy has crucial effects on electronic  structures of  $\mathrm{VSiGeN_4}$. Firstly, we consider that the magnetocrystalline direction  of  $\mathrm{VSiGeN_4}$ is along  out-of-plane.
At some representative $a/a_0$ values,  the energy band structures  with GGA+$U$+SOC  are plotted in FIG.7 of ESI, and
the evolutions of total energy band gap along with those at -K/K point  vs $a/a_0$ are shown  in \autoref{s-gap}.
Calculated results show that there are two points around  about $a/a_0$$=$0.9925  and 0.9945, where the total energy band gap is closed.
At the two strain points, conduction electrons are intrinsically 100\% valley polarized, and the HVM  state can be realized\cite{q10}.
At  about $a/a_0$$=$0.9925,  the band gap of K valley  gets closed, while a  band gap at -K valley can be observed.
At  about $a/a_0$$=$0.9945,  the band gap at -K valley is zero, while the band gap of K valley is kept.
The considered  strain ($a/a_0$) region can be divided into three parts by two HVM electronic states.

It is found that K and -K valleys of  both  valence and conduction bands are primarily contributed by the $d_{x^2-y^2}$+$d_{xy}$ or $d_{z^2}$  orbitals of V atoms, and the orbital characters energy band structures at representative  $a/a_0$$=$0.97, 0.993  and 1.02 from three regions are plotted in \autoref{s-sb}.
For 0.96$<$$a/a_0$$<$0.9925,  the $d_{x^2-y^2}$+$d_{xy}$ orbitals dominate K and -K valleys of  valence  bands, while the two valleys of  conduction bands  are mainly from $d_{z^2}$ orbitals (For example $a/a_0$$=$0.97). When $a/a_0$ is between 0.9925  and 0.9945, the $d_{x^2-y^2}$+$d_{xy}$/$d_{z^2}$  orbitals dominate K  valleys of  conduction/vallence  bands,  while orbital characters of  -K valley remain unchanged (For example $a/a_0$$=$0.993). For 0.9945$<$$a/a_0$$<$1.04, the distributions of  $d_{x^2-y^2}$+$d_{xy}$ and  $d_{z^2}$ orbitals  are opposite to ones of 0.96$<$$a/a_0$$<$0.9925 (For example $a/a_0$$=$1.02). These mean that there are two-time band inversion  between $d_{xy}$+$d_{x^2-y^2}$ and $d_{z^2}$ orbitals  with increasing $a/a_0$. The first occurs at K valley, accompanied by the first HVM state.
The second band inversion occurs at -K valley, along with the second HVM state.

The two HVM states imply  that  the total gap of $\mathrm{VSiGeN_4}$ closes  and reopens  two times, which suggests topological phase transition along with band inversion  between $d_{xy}$+$d_{x^2-y^2}$ and $d_{z^2}$ orbitals. The QAH state may appear, when   $a/a_0$ is between 0.9925  and 0.9945.  The  edge states at representative  $a/a_0$$=$0.993 are calculated to confirm  QAH  phase, which is plotted in \autoref{s-s}. It is clearly seen that a nontrivial chiral edge state connects  the conduction bands and valence  bands, implying a QAH phase. The calculated  Chern number $C$=-1,  which is also obtained
by integrating the Berry curvature (see \autoref{s-sb}) within the first BZ.
There are no nontrivial chiral edge states for the other two regions (0.96$<$$a/a_0$$<$0.9925 and 0.9945$<$$a/a_0$$<$1.04).
With increasing $a/a_0$,  two-time  topological phase transitions can be observed in monolayer $\mathrm{VSiGeN_4}$.
\begin{figure}
  \includegraphics[width=7cm]{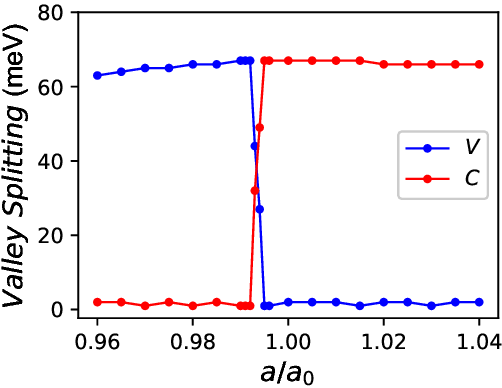}\\
\caption{(Color online)For $\mathrm{VSiGeN_4}$ monolayer with out-of-plane  magnetic anisotropy,  the absolute value of valley splitting   in both conduction (C) and valence (V) bands as a function of $a/a_0$.}\label{s-v}
\end{figure}

The transformations of  Berry curvatures  of K and -K valleys are related with these  topological phase transitions, and the distributions of Berry curvature are plotted in \autoref{s-sb} at representative  $a/a_0=$0.97, 0.993 and 1.02.
For 0.96$<$$a/a_0$$<$0.9925 and 0.9945$<$$a/a_0$$<$1.04, the Berry curvatures around -K and K valleys  have  the opposite
signs and different magnitudes. However,  for 0.9925$<$$a/a_0$$<$0.9945, the same signs and different magnitudes can be observed for Berry curvatures  around -K and K valleys. When $a/a_0$ changes from 0.96 to 1.04, there are twice topological phase transitions, which are related the flipping  of the sign of Berry curvature  at -K or K valley. For the first topological phase transitions,    the positive
  Berry curvature  ($a/a_0$$=$0.97)  changes into negative  one  ($a/a_0$$=$0.993) at K valley.
The second  topological phase transition is related with the sign flipping  of Berry curvature  of -K valley, and the negative  Berry curvature ($a/a_0$$=$0.993) changes into positive one  ($a/a_0$$=$1.02). These suggest  that strain can induce  sign-reversible  Berry curvature at K or -K valley, and this is  relevant to  topological phase transition.

\begin{figure}
  \includegraphics[width=8cm]{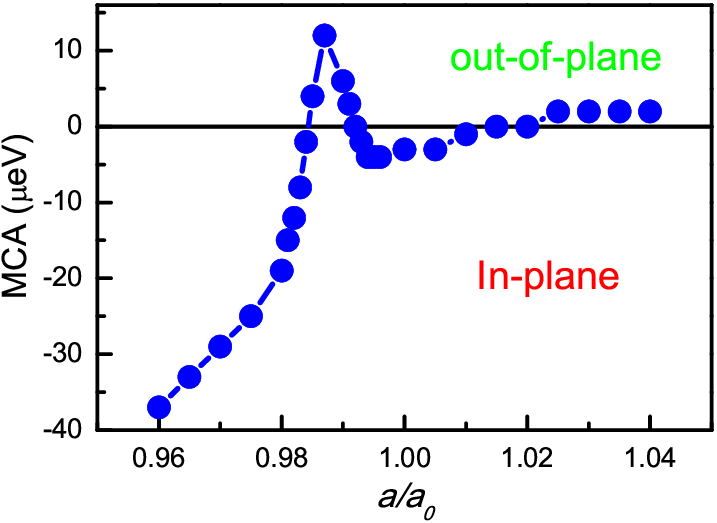}
   \includegraphics[width=8cm]{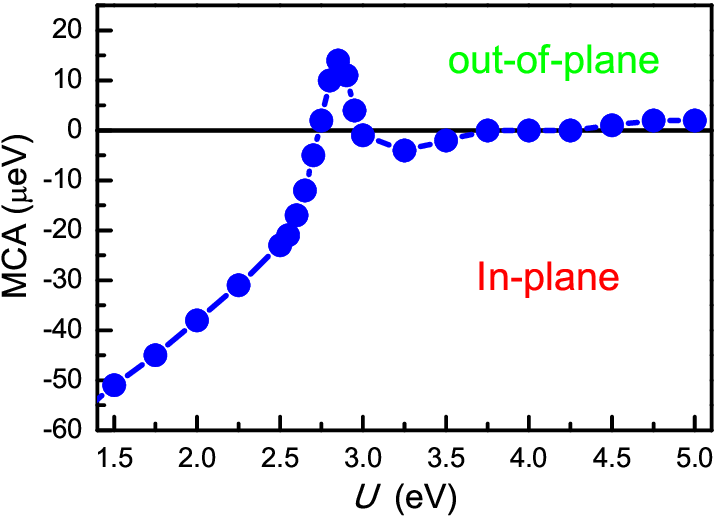}
\caption{(Color online)For $\mathrm{VSiGeN_4}$ monolayer,  the MCA energy  as a function of   $a/a_0$ (Top plane) and $U$ (Bottom plane).}\label{s-m}
\end{figure}

Calculated results show that $\mathrm{VSiGeN_4}$ monolayer has spontaneous valley polarization, and  the valley splitting for both valence and condition bands is plotted in \autoref{s-v}. For 0.96$<$$a/a_0$$<$0.9925,  the valley splitting of valence band is noteworthy, while the valley splitting of conduction band is very small.
However, for 0.9945$<$$a/a_0$$<$1.04, the opposite situation can be observed for valley splitting, compared with the case of 0.96$<$$a/a_0$$<$0.9925.
These can be explained by the distributions of  $d_{x^2-y^2}$+$d_{xy}$ and  $d_{z^2}$ orbitals (see \autoref{s-sb}).  If $d_{x^2-y^2}$+$d_{xy}$/$d_{z^2}$ orbitals  dominate  -K and K valleys, the valley splitting will be large/small. For 0.9925$<$$a/a_0$$<$0.9945, the valley splitting for both valence and condition bands is observable.  In this region, $\mathrm{VSiGeN_4}$  is a VQAHI with   spontaneous valley splitting and
 chiral edge states. For VAQHI, the  edge state has  a  special behavior of chiral-spin-valley  locking.
 For example $a/a_0$$=$0.993,   the edge state  in  \autoref{s-s} is spin up
with 100\% spin polarization and 100\% valley polarization, which is because the bands near the Fermi level  are  dominated by spin-up bands .
The edge state only appears at the K valley  due to the flipping of the sign of the Berry curvature  or band inversion at K valley. When the magnetization
is reversed, the edge state will move  to the -K valley with an opposite spin direction and chiral.

In quick succession,  we suppose  the magnetocrystalline direction  of
 $\mathrm{VSiGeN_4}$ monolayer along  in-plane one.
The energy band gaps as a function of  $a/a_0$ are plotted  in \autoref{s-gap}, and the representative energy band structures  are shown  in FIG.8 of ESI.
When $a/a_0$ changes from 0.96 to 1.04,  the gap firstly decreases, and  then increases. The corresponding $a/a_0$  of gap close is about 0.993.
It is found that no spontaneous valley polarization  in both  valence and conduction bands can be observed, and no QAH phase can be induced by strain.
When $a/a_0$ is less than about1.01 except 0.993 (semimetal), $\mathrm{VSiGeN_4}$ is a direct
gap semiconductor with VBM (CBM) at K/-K point. When $a/a_0$$>$1.01,  $\mathrm{VSiGeN_4}$ is s an indirect
gap semiconductor with CBM  at
the K/-K points, and the  VBM
 deviates  slightly from $\Gamma$ point. In short,  $\mathrm{VSiGeN_4}$ monolayer is a common FM semiconductor or semimetal.

Finally, we investigate the strain effects on MAE of $\mathrm{VSiGeN_4}$. We plot the MCA energy  as a function of $a/a_0$ in \autoref{s-m}. Strain-driven complex MCA (multiple transitions in the MCA) can be observed. In considered strain range, the MSA energy  changes from -19 $\mathrm{\mu eV}$  to -17 $\mathrm{\mu eV}$ to -15 $\mathrm{\mu eV}$, when the $a/a_0$ changes from 0.96 to 1.00 to 1.04. Calculated results show that the MAE is always negative within considered strain range, which means that strained $\mathrm{VSiGeN_4}$  is intrinsically common magnetic semiconductor.
However, the
magnetization can be adjusted from the in-plane to off-plane
direction through overcoming a small energy barrier  by the external magnetic field,
which will produce  valley
polarization and QAH phase. Within considered strain range, the largest  energy barrier (56 $\mathrm{\mu eV}$ at $a/a_0$=0.96) is equivalent to applying
a  external magnetic field of around 0.28-0.56 T.

\section{Discussion and Conclusion}
Monolayer $\mathrm{RuBr_2}$ shows  the same electronic states induced by strain  with $\mathrm{VSiGeN_4}$ for both out-of-plane and in-plane cases\cite{a7}.
However, the MCA energy of  $\mathrm{RuBr_2}$ varies  monotonously with increasing $a/a_0$. The strain can  suppress/enhance the
kinetic energy of electron, and then effectively
enhances/suppresses the correlation effect\cite{a5}.  This means that electronic correlation can induce the similar change of electronic states and MCA energy with strain, which has been confirmed for $\mathrm{RuBr_2}$\cite{a7}. To further confirm complex strain dependence of MCA energy, we calculate the MCA energy as a function of correlation strength $U$, which is also plotted in \autoref{s-m}. It is clearly seen that the MCA vs $a/a_0$ and MCA vs $U$ show very  similar behavior, as  expected.
The complex strain dependence of MCA  can be readily extended to $\mathrm{VSi_2P_4}$, $\mathrm{VSi_2N_4}$, $\mathrm{VSiSnN_4}$ and so on, because they share the same crystal structure with $\mathrm{VSiGeN_4}$. In fact, for $\mathrm{VSi_2P_4}$,  one
can observe similar transitions in the MCA energy as a function of $U$\cite{a5}.

In summary, we have demonstrated that strain  can result in a different
phase diagram for different magnetic anisotropy (out-of-plane and in-plane cases). For out-of-plane situation, the strain can induce
novel  VQAHI with exotic chiral-spin-valley locking edge states between two HVM states, and these are related with sign-reversible  Berry curvature and band inversions of $d_{xy}$+$d_{x^2-y^2}$ and $d_{z^2}$ orbitals at -K and K valleys. For in-plane situation, $\mathrm{VSiGeN_4}$ is a common magnetic semiconductor without spontaneous valley polarization.
Particularly, the calculated intrinsic MCA energy shows  multiple transitions  induced by strain, which is further confirmed by calculating MCA vs $U$.  Intrinsically, there is not  a VQAHI, which can be realized by external magnetic field.
 Our works deepen our understanding of strain  effects in the V-based 2D $\mathrm{MA_2Z_4}$ family materials,
and  open new perspectives for multifunctional electronic device applications based on these
materials.

\begin{acknowledgments}
This work is supported by Natural Science Basis Research Plan in Shaanxi Province of China  (2021JM-456),  Graduate Innovation Fund Project in Xi'an University of Posts and Telecommunications (CXJJDL2021001). We are grateful to Shanxi Supercomputing Center of China, and the calculations were performed on TianHe-2.
\end{acknowledgments}


\begin{references}

\bibitem{a1}N. D. Mermin and H. Wagner,  Phys. Rev. Lett. \textbf{17}, 1133 (1966).

\bibitem{a2} C. Gong, L. Li, Z. Li, H. Ji, A. Stern, Y. Xia, T. Cao, W. Bao,
C. Wang, Y. Wang, Z. Q. Qiu, R. J. Cava, S. G. Louie, J. Xia
and X. Zhang, Nature \textbf{546}, 265 (2017).


\bibitem{a3}B. Huang, G. Clark, E. Navarro-Moratalla, D. R. Klein, R. Cheng,
K. L. Seyler, D. Zhong, E. Schmidgall, M. A. McGuire, D. H.
Cobden, W. Yao, D. Xiao, P. Jarillo-Herrero and X. Xu, Nature \textbf{546}, 270 (2017).



\bibitem{a4}X. Liu, H. C. Hsu, and C. X. Liu, Phys. Rev. Lett. \textbf{111}, 086802 (2013).


\bibitem{a5}S. Li, Q. Q. Wang, C. M. Zhang, P. Guo and S. A. Yang,  Phys. Rev. B  \textbf{104}, 085149 (2021).


\bibitem{a6}S. D. Guo, J. X. Zhu, M. Y. Yin and B. G. Liu, Phys. Rev. B  \textbf{105}, 104416 (2022).

\bibitem{a7}S. D. Guo, W. Q. Mu and B. G. Liu, 2D Mater. \textbf{9}, 035011 (2022).



\bibitem{a8}S. Yang, Y.  Chen  and C. Jiang, InfoMat. \textbf{3}, 397 (2021).

 \bibitem{a9}Y. L. Wang and  Y. Ding, Appl. Phys. Lett. \textbf{119}, 193101 (2021).

\bibitem{a10}H. Huan, Y. Xue,  B. Zhao,  G. Y. Gao, H. R. Bao  and Z. Q. Yang, Phys. Rev. B \textbf{104}, 165427 (2021).


\bibitem{a11}Y. L. Hong, Z. B.  Liu, L. Wang  T. Y. Zhou,  W. Ma, C. Xu, S. Feng,
L. Chen, M. L. Chen, D. M. Sun, X. Q. Chen, H. M. Cheng and W. C. Ren, Science  \textbf{369}, 670 (2020).

\bibitem{a12}L. Wang, Y. Shi, M. Liu, A. Zhang, Y.-L. Hong, R. Li,
Q. Gao, M. Chen, W. Ren, H.-M. Cheng, Y. Li, and X.-
Q. Chen, Nature Communications \textbf{12}, 2361 (2021).

\bibitem{a12-1}Q. Wang, L. Cao, S. J. Liang et al.,  npj 2D Mater. Appl. \textbf{5}, 71 (2021).


\bibitem{a12-2}L. Cao,  G. Zhou, Q. Wang  L. K. Ang and  Y. S. Ang, Appl. Phys. Lett.  \textbf{118}, 013106 (2021).


\bibitem{a13}S. D. Guo, W. Q. Mu, Y. T. Zhu, R. Y. Han and W. C. Ren, J. Mater. Chem. C \textbf{9}, 2464 (2021).


\bibitem{a14}S. D. Guo, Y. T. Zhu, W. Q. Mu and X. Q. Chen, J. Mater. Chem. C \textbf{9}, 7465 (2021).



\bibitem{a15}D. Dey, A. Ray and L. P. Yu, arXiv:2203.11605 (2022).

\bibitem{1}P. Hohenberg and W. Kohn, Phys. Rev. \textbf{136},
B864 (1964); W. Kohn and L. J. Sham, Phys. Rev. \textbf{140},
A1133 (1965).



\bibitem{pv1} G. Kresse, J. Non-Cryst. Solids \textbf{193}, 222 (1995).

\bibitem{pv2} G. Kresse and J. Furthm$\ddot{u}$ller, Comput. Mater. Sci. 6, \textbf{15} (1996).

\bibitem{pv3} G. Kresse and D. Joubert, Phys. Rev. B \textbf{59}, 1758 (1999).
\bibitem{pbe}J. P. Perdew, K. Burke and M. Ernzerhof, Phys. Rev. Lett. \textbf{77}, 3865 (1996).


\bibitem{u}S. L. Dudarev, G. A. Botton, S. Y. Savrasov, C. J. Humphreys and A. P. Sutton, Phys. Rev. B \textbf{57}, 1505 (1998).


\bibitem{pv5}A. Togo, F. Oba, and I. Tanaka, Phys. Rev. B \textbf{78}, 134106
(2008).

\bibitem{bm}T. Fukui, Y. Hatsugai and H. Suzuki,  J. Phys. Soc. Japan. \textbf{74},
1674 (2005).


\bibitem{bm1}H. J. Kim,  https://github.com/Infant83/VASPBERRY, (2018).
\bibitem{bm2}H. J. Kim, C. Li, J. Feng, J.-H. Cho, and Z. Zhang, Phys. Rev. B  \textbf{93}, 041404(R) (2016).

\bibitem{w1}A. A. Mostofia, J. R. Yatesb, G. Pizzif, Y.-S. Lee, I. Souzad, D.
Vanderbilte and N. Marzarif,  Comput. Phys. Commun. \textbf{185}, 2309 (2014).

\bibitem{w2}Q. Wu, S. Zhang, H. F. Song, M. Troyer and A. A. Soluyanov, Comput. Phys. Commun. \textbf{224}, 405
(2018).

\bibitem{re1}X. B. Lu, R. X. Fei, L. H. Zhu and  L. Yang, Nat. Commun. \textbf{11}, 4724 (2020).

\bibitem{re2}M. Akram, H. LaBollita, D. Dey, J. Kapeghian, O. Erten and A. S. Botana, Nano Lett. \textbf{21}, 6633 (2021).

\bibitem{re3}H. J. F. Jansen,  Phys. Rev. B \textbf{59}, 4699 (1999).


\bibitem{re4}F. Xue,  Y. Hou, Z. Wang and R.  Wu, Phys. Rev. B  \textbf{100}, 224429 (2019).

\bibitem{re5}K. Sheng, Q. Chen , H. K. Yuan  and Z. Y. Wang, Phys. Rev. B \textbf{105}, 075304 (2022).


\bibitem{re5-1}J. L. Lado and J. Fern$\acute{a}$ndez-Rossier, 2D Mater. \textbf{4}, 035002 (2017).

 \bibitem{re6}P. Jiang, L. Kang, Y.-L. Li, X. Zheng, Z. Zeng, and S. Sanvito,
Phys. Rev. B \textbf{104}, 035430 (2021).

\bibitem{re7}S. Zhang, R. Xu, W. Duan, and X. Zou, Adv. Funct. Mater. \textbf{29},
1808380 (2019).

\bibitem{ela}E. Cadelano and L. Colombo, Phys. Rev. B  \textbf{85}, 245434 (2012).
\bibitem{ela1}Y. X.Wu,  W. Sun, S. Y. Liu,  B. Wang,  C. Liu, H. B. Yin and Z. X. Cheng, Nanoscale  \textbf{13}, 16564 (2021).

\bibitem{f6}W. Y. Tong, S. J. Gong, X. Wan, and C. G. Duan,
Nat. Commun. \textbf{7}, 13612 (2016).

\bibitem{v2}P. Zhao, Y. Dai, H. Wang, B. B. Huang and  Y. D. Ma, ChemPhysMater, \textbf{1}, 56 (2022).

\bibitem{v3}R. Li, J. W. Jiang, W. B. Mi  and H. L. Bai, Nanoscale  \textbf{13}, 14807 (2021).


\bibitem{q9}D. Xiao, M. C. Chang, and Q. Niu, Rev. Mod. Phys. \textbf{82}, 1959
(2010).

\bibitem{q10} H. Hu, W. Y. Tong, Y. H. Shen, X. Wan, and C. G. Duan, npj
Comput. Mater. \textbf{6}, 129 (2020).

\end{references}
\end{document}